\documentclass[doublecol]{epl2} 
\usepackage{graphicx}% Include figure files
\usepackage{bm}% bold math
\usepackage[english]{babel}
\usepackage[latin1]{inputenc} 
\usepackage{graphicx}
\usepackage{epstopdf}
\usepackage{amsfonts}
\usepackage{color}
\usepackage{epsfig}
\usepackage{mathrsfs}
\usepackage{amsmath,amssymb}

\renewcommand{\(}{\left(}
\renewcommand{\)}{\right)}

\title{Fast detection of nonlinearity and nonstationarity in short and noisy time series}
\shorttitle{Fast detection of nonlinearity in short and noisy time series}
\author{M. De Domenico\inst{1,2} \and V. Latora\inst{1,2}}

\institute{                    
  \inst{1} Laboratorio sui Sistemi Complessi, Scuola Superiore di Catania - Via San Nullo 5/i, 95123 Catania, Italy\\
  \inst{2} Dipartimento di Fisica e
    Astronomia, Universit\`a di Catania, and INFN - Via S. Sofia 64, 95123
    Catania, Italy
}
\abstract{  We introduce a statistical method to detect nonlinearity and
  nonstationarity in time series, that works even for short sequences
  and in presence of noise.  The method has a discrimination power
  similar to that of the most advanced estimators on the market, yet
  it depends only on one parameter, is easier to implement and faster.
  Applications to real data sets reject the null hypothesis of an
  underlying stationary linear stochastic process with a higher
  confidence interval than the best known nonlinear discriminators up
  to date.}

\begin{document}

%\graphicspath{{figure/}}
\selectlanguage{english}

%\preprint{\emph{SSC, Laboratory for Complex Systems }, DRAFT \today}

\maketitle

%%%%%%%%%%%%%%%%%%%%%%%%%%%%%%%%%%%%%%%%%
%\section{Introduction}

  Natural phenomena are often studied through measures of physical
  observables that change over time. Hence, time series analysis is of
  extraordinary importance for the comprehension and the
  characterization of a physical process. The analysis of a time
  series should generally be able to detect, within some confidence
  level, the stochastic or deterministic nature of the underlying
  process, and eventually to quantify the degrees of freedom, the
  presence of nonlinearity or nonstationarity, and finally the
  predicability of the future states.
  Experimental time series are affected by measurement error, and they
  are often corrupted by unknown noise sources. In addition to this,
  while some processes, such as laser emissions \cite{Huebner89} or
  network traffic \cite{Meloni08}, can produce a large amount of data
  in few hours, other natural processes, such as sunspots
  \cite{Palus99,Timmer00,Palus00} or seismic events \cite{Mega03}, may
  require long times of observation to obtain relevant 
  informations. As a consequence, methods of time series analysis
  should be able to work on both short and long noisy series.

  The surrogate data method provides a rigorous statistical approach
  to the nonlinear features detection of a time series
  \cite{Theiler92,Prichard94}. The method consists in formulating a
  null hypothesis, e.g. ``the time series is generated by a linear and
  stationary stochastic process'', an alternative hypothesis,
  e.g. ``the time series is \emph{not} generated by a stationary
  linear stochastic process'', and in preparing a set of $N$
  constrained surrogate time series, generally with the same linear
  statistical features as the original one. One or more observables,
  so-called {\em estimators}, {\em discriminators}, or {\em
    discriminating statistics}, are obtained from both the original
  time series and for the surrogates representing the null hypothesis:
  by performing a nonparametric test, as the rank order, observables
  are statistically analysed and the rejection of the null hypothesis
  is claimed within a certain confidence level. As pointed out in
  Refs.~\cite{Theiler92,Prichard94}, a large number of different
  measures have been considered over the years to detect nonlinearity
  in time series: higher order statistics, time reversal asymmetry,
  correlation dimension \cite{Schmitz97}, largest Lyapunov exponent,
  nonlinear prediction error (NPE) \cite{Farmer87, Sugihara90,
    Schmitz97} and Volterra-Wiener-Korenberg polynomials
  \cite{Barahona96}, just to mention those most commonly adopted.
  However, some of the measures either have low discrimination power
  \cite{Schmitz97}, or are not able to distinguish chaos from coloured
  noise, as the correlation dimension \cite{Provenzale89,
    Provenzale91, Provenzale92}, or require long time series,
  generally not available in the real world. In particular, the NPE, a
  method based on phase-space reconstruction and that requires three
  parameters, has the highest discrimination power on short and noisy
  time series, and gives either better or comparable performance than
  other methods \cite{Schmitz97}.

\bigskip
In this Letter, we introduce a fast method to reject the null
hypothesis of an underlying stationary linear stochastic process.  
The method does not require any embedding procedure for phase 
space reconstruction, and is based on the evaluation of the 
differences between the original time series and the 
stationary linear model that best approximates it.   
The kurtosis of the distribution of differences is finally 
used as discrimination statistics to reject the null hypothesis.
The method works well with both deterministic and stochastic
series, either time continuous or discrete. It is simpler to compute
and faster than other excellent discriminators, and it appears to be
robust even for very short time series, highly corrupted with measurement
noise. Indeed, as an application to a still open question, we will
show how the method improves up to 98\% the confidence level for the
rejection of null hypotheses in the cases of the monthly
  smoothed sunspots index.

%%%%%%%%%%%%%%%%%%%%%%%%%%%%%%%%%%%%%%%%%
%\section{Method}

Given a $l$-samples univariate time series $\{s_{n}\}$, with
$n=1,2,\ldots,l$, we produce a set $\{\hat{s}_{n}^{(i)}\}$,
$i=1,2,...,N$, of $N$ surrogates of $\{s_{n}\}$. Each of the surrogate
is a time series having the same mean, variance, density function,
power spectrum and, thus, autocorrelation function as the original
time series $\{s_{n}\}$. Different procedures to produce surrogates
fulfilling several requirements have been introduced over the
years \cite{Schreiber98,Schr-Schm00}.  In particular, here, we adopt
an iterative amplitude adjusting Fourier transform (IAAFT) scheme
\cite{Schr-Schm96, Schreiber98}.  Since the density function and the
power spectrum of $\{s_{n}\}$ and $\{\hat{s}_{n}^{(i)}\}$ are the
same, nonlinearity or nonstationarity, if present, should emerge from
the analysis of the deviations from the best stationary linear
  approximation of the time series. The simplest stationary
  linear model is the autoregressive AR($p$) of order $p$, defined as
\begin{eqnarray}
\label{ARp}
x_{n}=a_{0}+\sum_{i=1}^{p}a_{i}x_{n-i} + \epsilon_{n}
\end{eqnarray}
where the integer $p$ is the memory range of the model, and 
$\epsilon_{n}$ is a stochastic stationary process with
$\langle\epsilon\rangle=0$. 
To ensure the stationarity of the process,
the coefficients $a_{i}$ in Eq.~(\ref{ARp}) have to be chosen such
that the roots of the characteristic polynomials: $ 
1-\sum_{i=1}^{p}a_{i}z^{i}=0$ lie outside the unit circle. AR($p$) are
well known models used in time series fitting problems, and a wide
literature covers the exact and efficient estimation of the
coefficients $a_{i}$ (see Ref.~\cite{Brockwell09} for an extensive and 
up-to-date review). 
\begin{figure*}[t]
  \begin{center}
      \includegraphics[angle=0, scale=0.45]{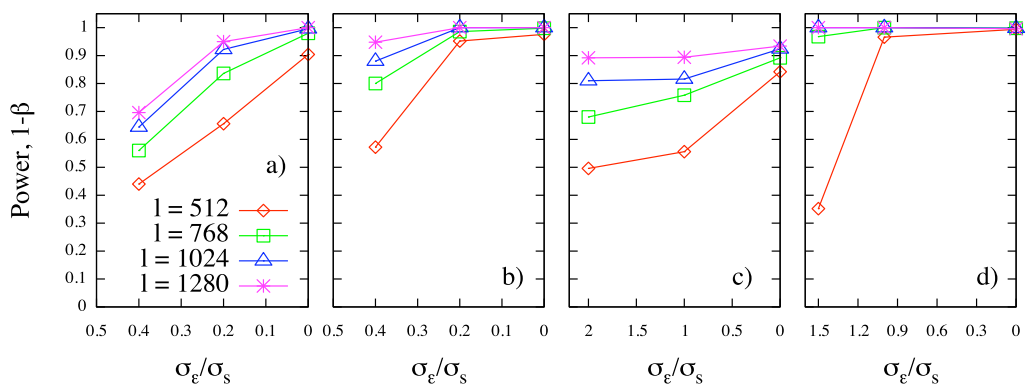}
      \caption{{ARK power versus noise strength for: a) Mackey-Glass
          map in the chaotic regime with the addition of correlated
          noise,b) Ikeda map in the chaotic regime, with uncorrelated
          noise, c) nonstationary autoregressive moving average (ARMA) with
          correlated noise, d) Lorenz model in chaotic regime with
          uncorrelated noise.}}
    \label{fig-powers}
  \end{center}
\end{figure*}
Here, we adopt a least squares approach, solved by a QR
  decomposition algorithm. As a result of fitting the original time
  series $\{s_{n}\}$ with an AR($p$), we get the coefficients 
$\tilde{a}_{i}$, the model $\{\tilde{x}_{n}\}$, and the
  residuals series $\{\tilde{\epsilon}_{n}\}$, defined as 
  $\tilde{\epsilon}_{k}=s_{k}-\tilde{x}_{k}$ for $k=1,2,...,n$.  As for 
the order $p$ of the autoregressive model that best approximates the
  original time series $\{s_{n}\}$, we have chosen the one that minimizes an
  information criterion, according to Akaike \cite{Akaike74} and
  Schwarz \cite{Schwarz78}. The estimator of linearity and
stationarity we propose is the kurtosis of the residuals:
\begin{eqnarray}
\mathcal{K}=\frac{\left\langle \(\tilde{\epsilon}-\langle\tilde{\epsilon}\rangle\)^{4} \right\rangle}{\left\langle \(\tilde{\epsilon}-\langle\tilde{\epsilon}\rangle\)^{2} \right\rangle^{2}}
\label{ark}
\end{eqnarray}
The same fitting procedure is repeated for each of the $N$ surrogate
series $\{\hat{s}_{n}^{(i)}\}$, and the obtained estimators are
respectively indicated as $\hat{\mathcal{K}}^{(i)}$. Finally, the test
against the null hypothesis is based on comparing $\mathcal{K}$ to the
distribution of $\hat{\mathcal{K}}^{(i)}$. We name the discriminator 
in Eq.~\ref{ark} 
\emph{autoregressive-fit residuals kurtosis} (ARK).

Summing up, the method consists, in practice, in the following steps: 

\begin{enumerate}
\item Given a time series $\{s_{n}\}$, obtain the best stationary
  linear model that minimizes an information criterion. 

\item Build the time series $\tilde{\epsilon}_{n}$ of residuals and 
evaluate the kurtosis of their density. 

\item Fix a significance level $\alpha$, and create $N= 2/ \alpha -1$ 
 surrogates of $\{s_{n}\}$. Repeat points 1 and 2 for each surrogate
  time series $\{\hat{s}^{(i)}_{n}\}$. 

\item Finally, compare the values of the discriminator in Eq.~(\ref{ark})   
obtained for $\{s_{n}\}$ and $\{\hat{s}_{n}^{(i)}\}$ with a two-sided rank order
  test of size $\alpha$.
\end{enumerate}

%%%%%%%%%%%%%%%%%%%%%%%%%%%%%%%%%%%%%%%%%
%\section{Results}

We have tested ARK with a wide variety of short time series, either
from models and from real data sets. 

\bigskip
~~~\textit{Synthetic time series.~~}
We first considered stationary linear autoregressive moving-average
models (ARMA), i.e. discrete null models. We performed an extensive and
systematic analysis of ARK statistics, by varying both autoregressive
and moving average orders. As expected, for a discriminating statistics 
evaluated on null models, we obtained a flat density of p-values. 
The same result was obtained when considering stationary linear
Langevin processes, i.e. continuous null models. It follows that ARK
is not biased against the null hypothesis. 

We then addressed autoregressive moving-average models in
  nonstationary regime, and several well known nonlinear models, both
  in chaotic and non chaotic regime. In particular, we tested the
  stability of the method in discriminating nonlinearity and
  nonstationarity under the contamination of stochastic noise,
  correlated or uncorrelated, of increasing strength, by calculating
  the probability $\beta$ to accept the null hypothesis when the
  alternative is true (also known as type II error). For each
  simulated time series with variance $\sigma_{s}^{2}$, we added
  artificial noise with variance $\sigma_{\epsilon}^{2}$, and we
  studied the so-called {\em power} $1-\beta$ of ARK by
  varying the ratio $\sigma_{\epsilon}/\sigma_{s}$. In practice, the
  results obtained where the method has a power of less than 70\% are
  considered questionable \cite{Schmitz97}.  In
  Fig. \ref{fig-powers} we report the power $1-\beta$ versus the
  relative noise level $\sigma_{\epsilon}/\sigma_{s}$, for different
  dynamical systems. The results were obtained with $10^4$ rank order
  tests of size $\alpha=1\%$ ($N=199$), corresponding to a confidence
  level of $99\%$. We considered both deterministic and stochastic
  series, either time continuous or discrete. In the case of
  noise-free nonlinear or nonstationary time series, we get a
  discrimination power close to 100\%. Even in the presence of noise,
  the ARK exhibits an excellent discrimination power. For all the
  considered dynamical systems, a sequence of $l=1024$ samples is
  enough to achieve a discrimination power larger than 70\%, even for
  relative noise levels as large as 40\% in the case of maps, and as
  large as 150\% in the case of the nonstationary ARMA and of the
  Lorenz system. Notice that adding a relative noise level of 150\% to
  a time series corresponds to superimposing a noise amplitude that is
  1.5 times the intrinsic fluctuations of the original series. Indeed,
  in the case of a chaotic dynamics, such as the Lorenz system
  reported in panel d), it is possible to reach a discrimination power
  close to 100\% also for shorter length ($l=768$). This result is
  particularly relevant if compared against the result from NPE: 
  in presence, for instance, of a relative noise level of 150\%, the phase space of the Lorenz model is not 
  well reconstructed through the delay embedding method, and, in fact, a power of no
  more than 30\% is achieved. Hence, the results in Fig. \ref{fig-powers} show that our procedure is suited for the analysis of real datasets, generally noisy ($\sigma_{\epsilon}/\sigma_{s}<1$) time series of
  few thousands of samples. 
  
\begin{figure}[h]
  \begin{center}
      \includegraphics[angle=0, scale=0.25]{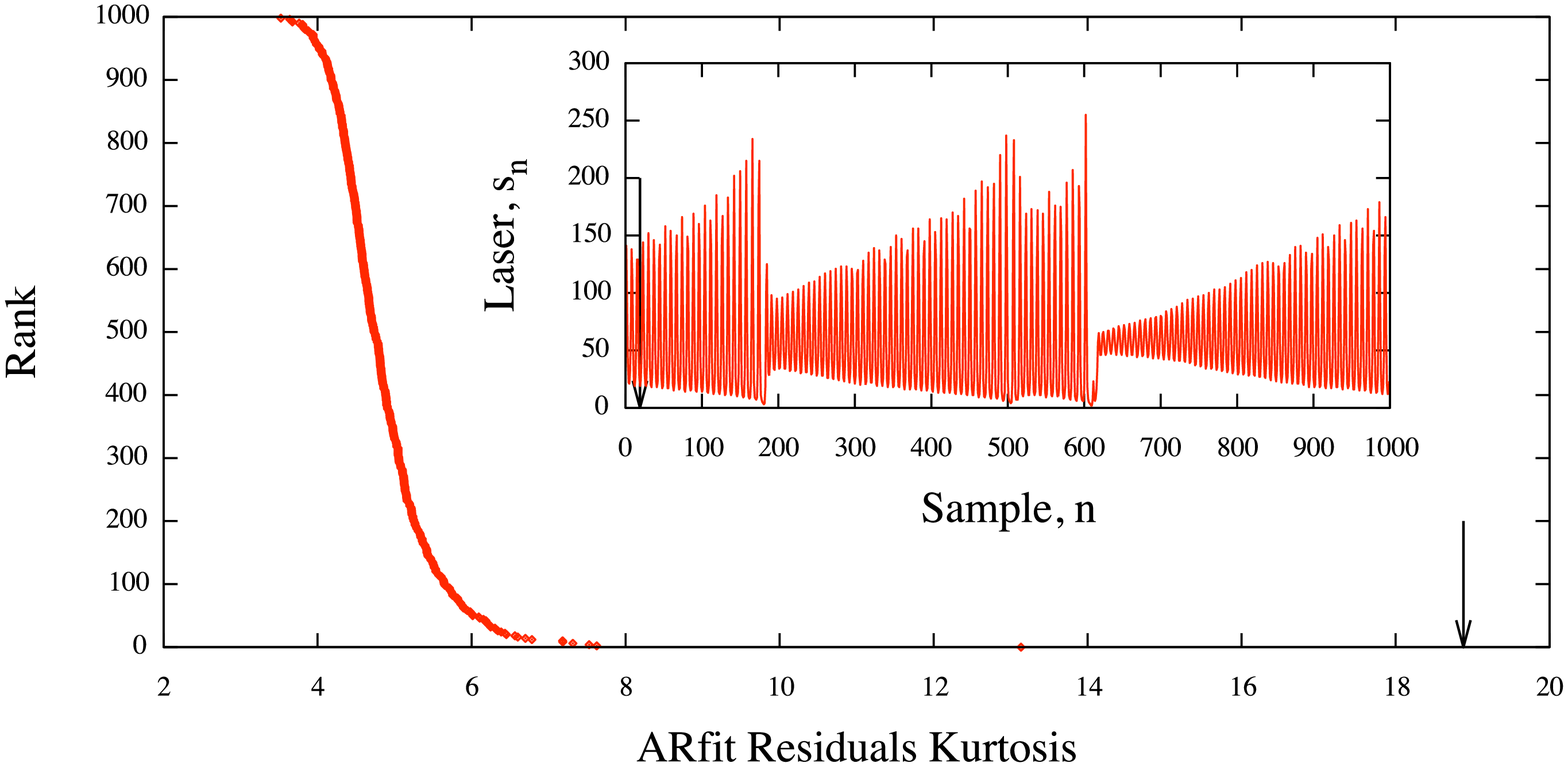}a)
      \includegraphics[angle=0, scale=0.25]{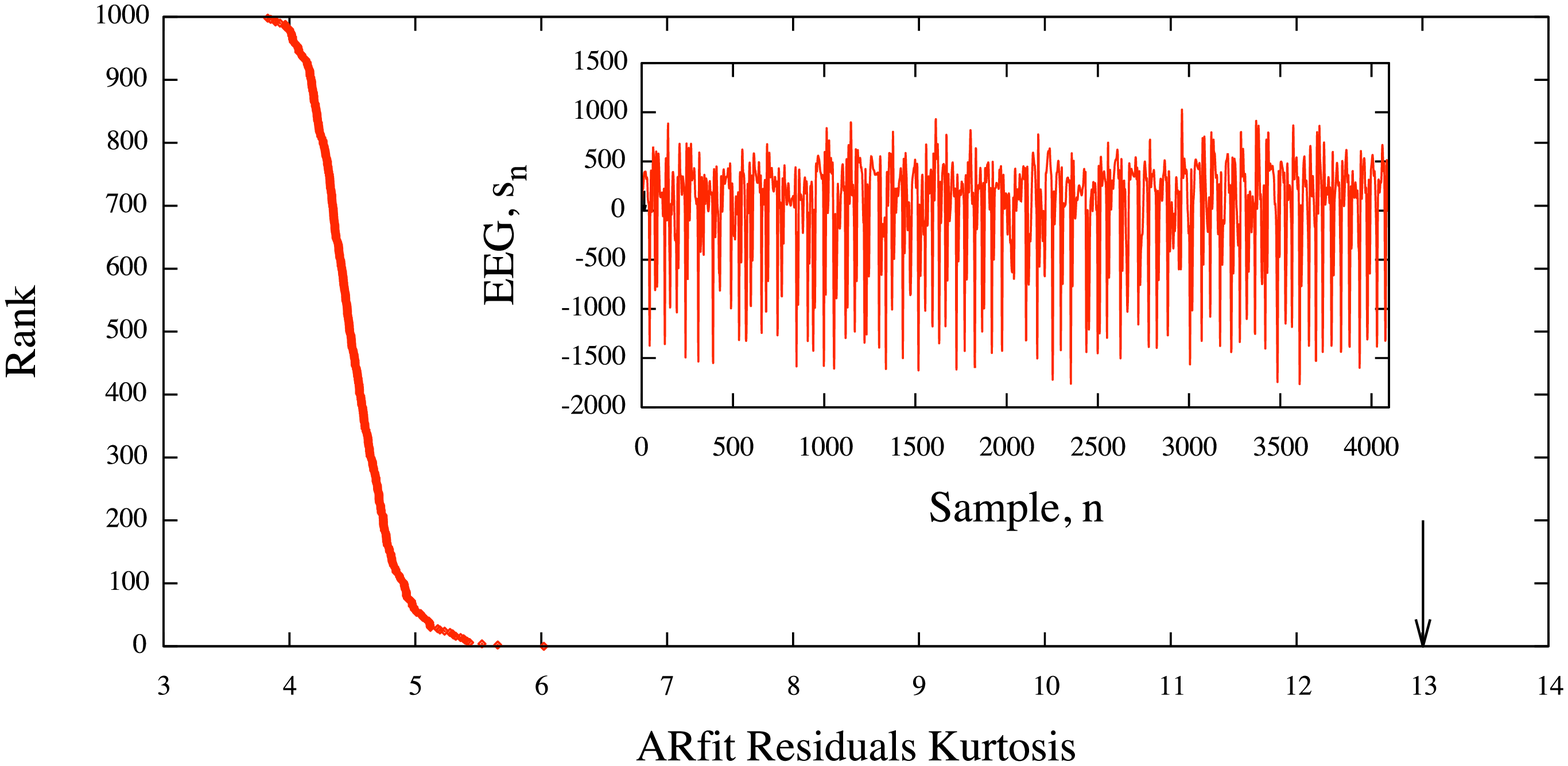}b)
    \caption{{Rank order test (1000 surrogates) for the Far Infrared Laser intensity time series (a) and for an EEG recording during epileptic seizure (b). The arrow indicates the value of ARK for the original time series, dots indicate the rank ordered values of ARK for their surrogates.}}
    \label{fig-laser}
  \end{center}
\end{figure}

\bigskip
~~~\textit{Real data sets.~~}
The first nonlinear experimental time series we have considered is a sample of
$l=1000$ values of the intensity of a Far Infrared Laser (FIR) in a
chaotic state (see Ref.~\cite{Huebner89}). In Fig. \ref{fig-laser}a we show the rank ordered ARK values for both the FIR time series and 1000 of its surrogates: as expected, the rank order test rejects the null hypothesis with 99.8\% confidence level.

The second nonlinear experimental time series we have considered is a sample of $l=4096$ values of intracranial EEG recordings during epileptic seizures (see Ref.~\cite{Andrzejak01}). In Fig. \ref{fig-laser}b we show the rank ordered ARK values for both the EEG time series and 1000 of its surrogates: as expected, the rank order test rejects the null hypothesis with 99.8\% confidence level.

As third application to real databases, the discrimination power of
ARK is tested against time series of sunspots index. In particular,  
we consider the monthly smoothed sunspots index from 1749 to 2008, 
and the 11-years averaged sunspots index reconstructed up to the past 
11,400 years. The sunspots index, introduced by Wolf in 1848, is strongly
related to the solar cycle discovered by Schwabe in 1843. Such series 
shows strong irregularities, partially explained
by magnetohydrodynamics of dynamo, and a quasi-periodic
behavior. However, the real nature of solar cycle is still debated, as
shown in the following. Gurbuz and Beck \cite{Beck95} claimed that the dynamics of successive
sunspot maxima is low-dimensional with features similar to the
intermittent logistic map. Barnes \cite{Barnes80} proposed an ARMA(2,2) model mapped by a nonlinear
function to reproduce the solar cycle with its statistical features
and its irregular and unpredictable behavior. In conflict with this result, a randomly driven nonlinear oscillator was proposed for the first time
by Palu\^s and Novotn\'a \cite{Palus99} by using the mutual dependence
of the instantaneous amplitude and frequency of sunspot series as
discriminator in a surrogate data test, to reject the null hypothesis
of an underlying Barnes model \cite{Barnes80}. The mapping to a scalar time series
from the spatio-temporal magnetic field described by a nonlinear,
eventually stochastically, driven partial differential equation for
the magnetohydrodynamic dynamo, was not excluded from the results
\cite{Timmer00, Palus00}.
Sunspot irregularities were attributed to the stochastic fluctuation
in one of the parameters of a Van Der Pol nonlinear oscillator
describing the irregular periodic magnetic field \cite{Mininni00},
while recently, an overembedding approach introduced for modeling and
prediction of nonstationary systems was successfully applied to the
series with high precision \cite{Verdes06}.

%\begin{figure}[!htb]
%  \begin{center}
%      \includegraphics[angle=0, scale=0.25]{figure/sunspot_millennium.eps}a)
 %      \includegraphics[angle=0, scale=0.25]{figure/barnes_millennium.eps}b)
 %   \caption{{Annual sunspots index, averaged on 11-years, reconstructed up to the past 11,400 years \cite{Solanki04} (a) and a corresponding Barnes model (b).}}
  %  \label{sunmillennium}
 % \end{center}
%\end{figure}

\begin{figure}[!htb]
  \begin{center}
      \includegraphics[angle=0, scale=0.25]{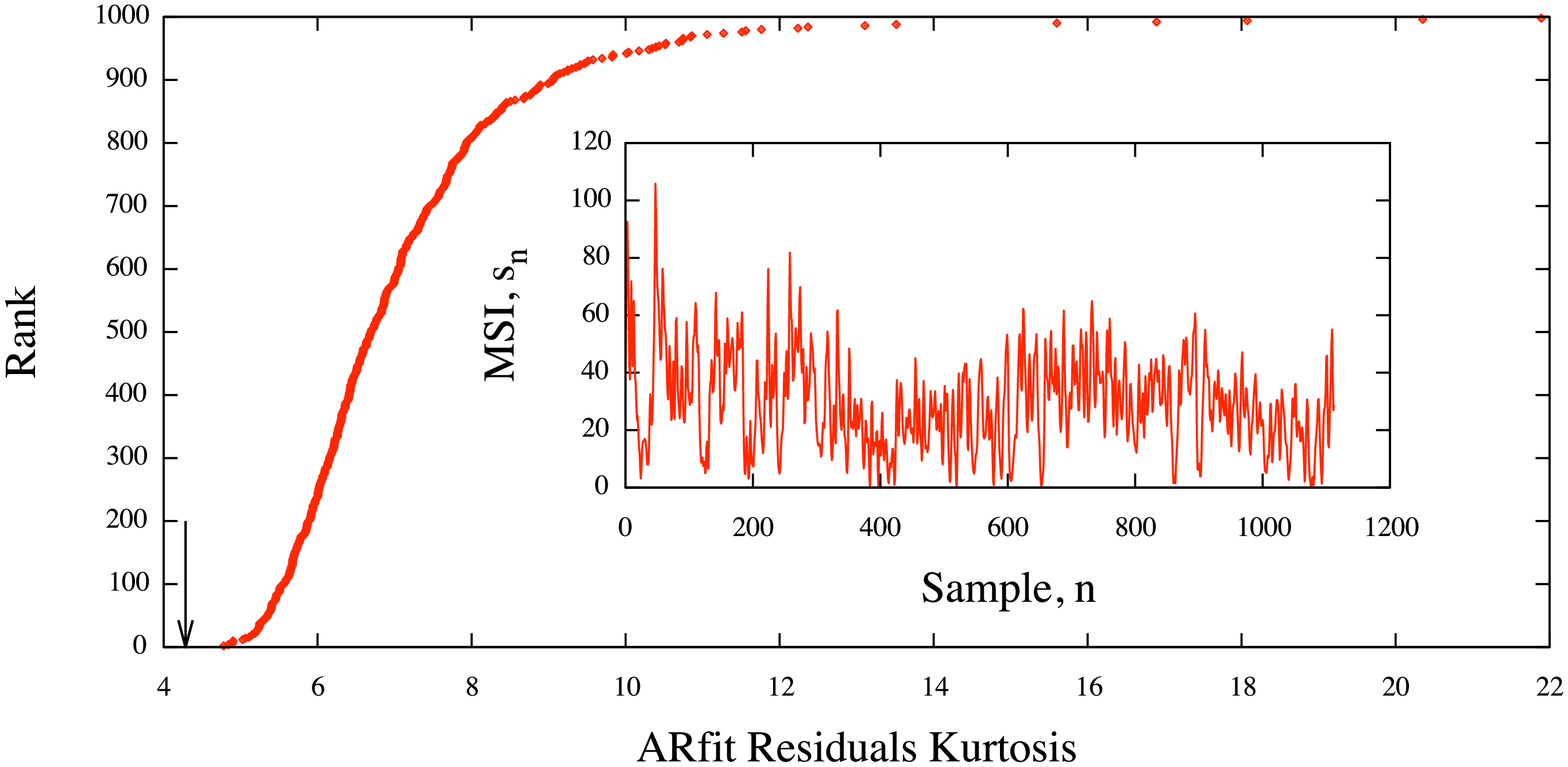}a)
      \includegraphics[angle=0, scale=0.25]{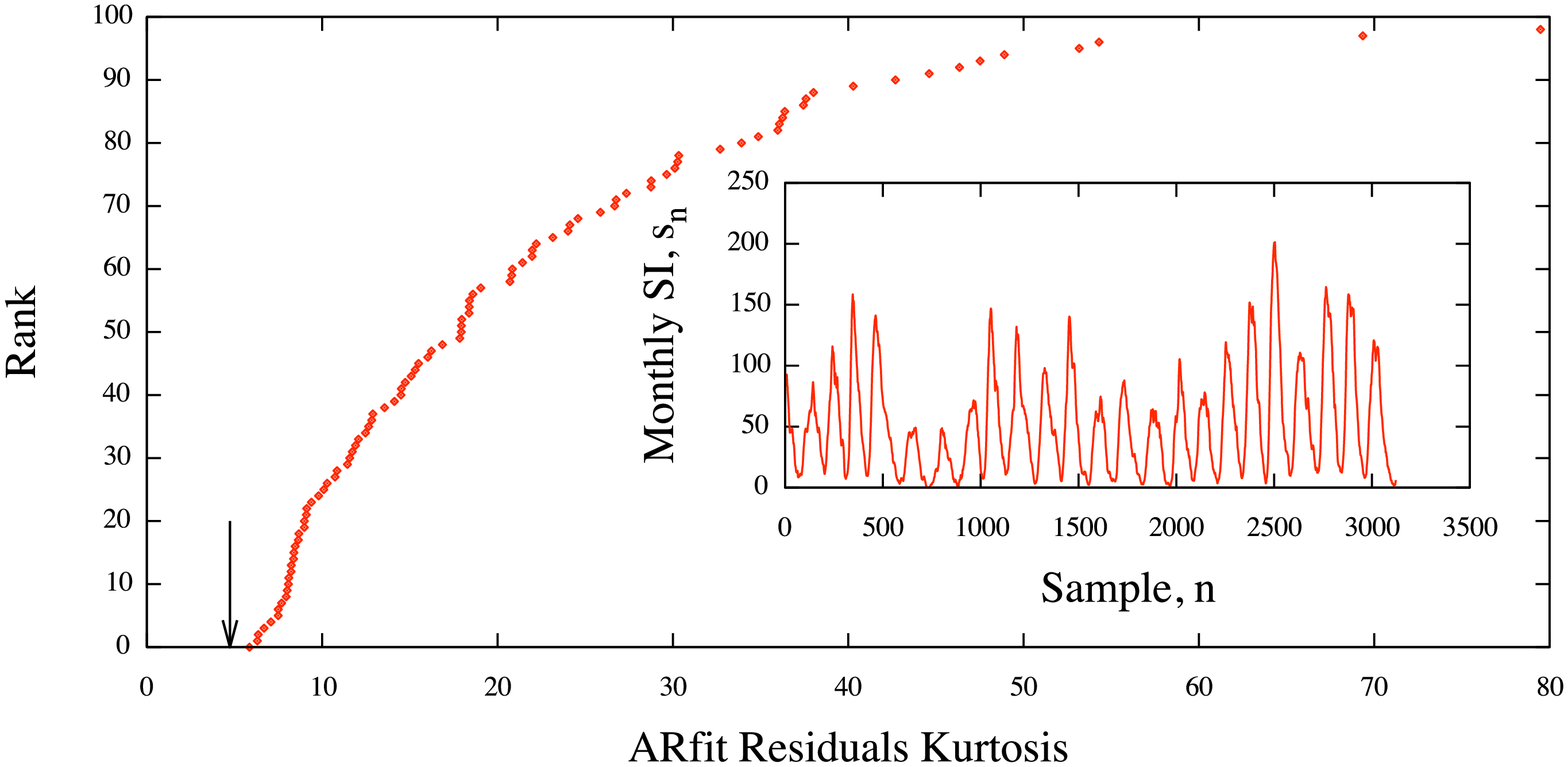}b)
    \caption{{Rank order test for the sunspots index,
averaged on 11-years, reconstructed up to the past 11,400 years against 1000 Barnes null models (a), and for the monthly smoothed sunpost index from 1749 to 2008 and 100 surrogates (b). The arrow indicates the value of ARK for the original time series, dots indicate the rank ordered values of ARK for their surrogates.}}
    \label{fig-msi}
  \end{center}
\end{figure}

We considered the series of $l=3123$ samples for the monthly smoothed sunspots index, released by the Solar Influences Data analysis Center \cite{SIDC} from 1749 to 2008, and the historical series of $l=1113$ samples for the sunspots index,
averaged over 11-years, reconstructed up to the past 11,400 years (MSI)
through indirect methods \cite{Solanki04,Usoskin03}. Standard estimators as the Lyapunov exponent, the correlation dimension,
and the increase of the prediction error with the prediction horizon
can lead to spurious results when applied to short time series, as
shown for instance in Ref.~\cite{Mininni00} and references therein. 

Here, we have applied ARK both to the $l=1113$ and to the $l=3123$ series, performing surrogate tests against
different null models. If a null hypothesis can not be rejected, the
sunspots index time series is not distinguishable from the null model:
thus either it is well described by the model or the used statistics has
not enough discrimination power. We started by considering the short time series and Barnes time series as null models. Fig. \ref{fig-msi}a shows the rank ordered ARK values for the historical series and 1000 of its nulls: we reject the null hypothesis of an underlying Barnes process with 99.8\% confidence level. Successively, the long time series was tested against IAAFT surrogates. Fig. \ref{fig-msi}b shows the rank ordered ARK values for the monthly series and 100 of its surrogates: we reject the null hypothesis of an underlying stationary linear stochastic process with 98\% confidence level.

\begin{figure}[!htb]
  \begin{center}
       \includegraphics[angle=0, scale=0.25]{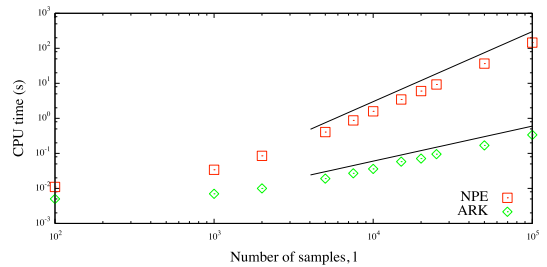}
    \caption{{CPU time in seconds versus the length of the series, for both NPE and ARK applied to Lorenz models of increasing number of samples.}}
    \label{bench}
  \end{center}
\end{figure}

We verified that one of the best test statistics up to date \cite{Schmitz97}, namely the nonlinear prediction error, gives similar results. Indeed, we emphasize that NPE makes use of three parameters, the lag time $\nu$, the embedding dimension 
$M$ and the size of neighbourhoods, while ARK does not need any phase space reconstruction, and only
uses one parameter. In addition to this, our numerical
experiments on simulated time series show that, excluding the time for the estimation of needed parameters, ARK complexity is generally $\mathcal{O}(l)$, while NPE complexity is $\mathcal{O}(l^{2})$ \cite{TISEAN}, where $l$ is the time series length. Fig. \ref{bench} shows the CPU time in seconds versus the length of the series, for both NPE and ARK applied to Lorenz models of increasing number of samples: complexity is $\mathcal{O}(l^{1.96\pm0.02})$ for the former and $\mathcal{O}(l^{0.96\pm 0.03})$ for the latter.

In this Letter, we have introduced ARK analysis, a new procedure to
detect nonlinearity and nonstationarity in a time series. We have 
tested the discrimination power of ARK on a wide variety of
short and noisy time series, either from models and from real data
sets. A series of the far infrared laser emission in a chaotic state and a sample of the intracranial EEG recordings during epileptic seizure, have been analyzed: in both case our procedure was able to detect, with high confidence level, the presence of nonlinearity.
 Using Barnes models, supposed to replicate the behavior of the annual
sunspots index, we have obtained a statistically significant rejection
of the null hypothesis of an underlying stationary linear stochastic
process, possibly mapped by a nonlinear function, for the historical time series. Using surrogates, we have obtained a statistically significant rejection of the null hypothesis of an underlying stationary linear stochastic process, for the monthly smoothed sunspot index time series. Although no particular model for sunspots index has been proposed here, the presented results are an
important step for the comprehension of the underlying solar cycle
mechanisms. We believe that our method can be successfully
applied to other interesting real-world time series whose underlying
dynamics is still debated.

%\acknowledgments
%Insert here the text.

%\begin{thebibliography}{0}

\bibliographystyle{eplbib} % stile della bibliografia
\bibliography{draft}

%\end{thebibliography}

\end{document}